\documentclass{elsart}
\usepackage{epsfig}
\newcommand{\g}{$\gamma$}

\newcommand{\cs}{$^{137}$Cs}

\newcommand{\na}{$^{22}$Na}
\newcommand{\co}{$^{60}$Co}
\newcommand{\yt}{$^{88}$Y}

\newcommand{\pot}{$^{40}$K}

\newcommand{\bc}{\begin{center}}
\newcommand{\ec}{\end{center}}
\newcommand{\be}{\begin{equation}}
\newcommand{\ee}{\end{equation}}
\newcommand{\bfg}{\begin{figure}}
\newcommand{\efg}{\end{figure}}
\newcommand{\bi}{\begin{itemize}}
\newcommand{\ei}{\end{itemize}}
\newcommand{\bt}{\begin{table}}
\newcommand{\enta}{\end{table}}
\newcommand{\keV}{\mbox{ke\hspace{-0.1em}V}}
\newcommand{\MeV}{\mbox{Me\hspace{-0.1em}V}}

\newcommand{\phibar}{\ensuremath{\bar{\varphi}}}

\newcommand{\phigeo}{\ensuremath{\varphi_\mathrm{geo}}}
\newcommand{\ephibar}{\ensuremath{\sigma_{\cos{\bar{\varphi}}}}}
\newcommand{\ephigeo}{\ensuremath{\sigma_{\cos{\varphi}_\mathrm{geo}}}}
\renewcommand{\deg}{\ensuremath{^\circ}}
\newcommand{\gcc}{\mbox{g cm$^{-3}$}}
 
\begin{document}

\begin{frontmatter}

\title{Compton Imaging of MeV Gamma-Rays with the Liquid Xenon
Gamma-Ray Imaging Telescope (LXeGRIT)}  
\author{E.~Aprile$^a$},
\author{A.~Curioni$^{a,}$\thanksref{yale}},
\author{K.L.~Giboni$^a$}, 
\author{M.~Kobayashi$^{a,}$\thanksref{waseda}},
\author{U.G.~Oberlack$^b$},
\author{S.~Zhang$^{a,}$\thanksref{sz}}   
\address{$^a$ Columbia Astrophysics Laboratory and Physics Department,
Columbia University, New York, NY, USA}  
\address{$^b$Dept. of Physics \& Astronomy, Rice University, Houston,
TX, USA}  
\thanks[yale]{Present address: Yale University, Physics Dept., New Haven, CT, USA}
\thanks[waseda]{Present address: Waseda University,Tokyo, Japan}
\thanks[sz]{Present address: High Energy Astrophysics Lab, Institute
of High Energy Physics, Beijing, China}

\begin{abstract}

The Liquid Xenon Gamma-Ray Imaging Telescope (LXeGRIT) is the first
realization of a liquid xenon time projection chamber for Compton
imaging of MeV \g-ray sources in astrophysics. By measuring the
energy deposit and the three spatial coordinates of individual
\g-ray scattering points, the location of the source in the sky is
inferred with Compton kinematics reconstruction. The angular
resolution is determined by the detector's energy and spatial
resolutions, as well as by the separation in space between the first
and second scattering. The imaging response of LXeGRIT was established
with \g-rays from radioactive sources, during calibration and
integration at the Columbia Astrophysics Laboratory, prior to the 2000
balloon flight mission. In this paper we describe in detail the
various steps involved in imaging  sources with LXeGRIT and present 
experimental results on angular resolution and other parameters which
characterize its performance as a Compton telescope. 

\end{abstract}

\end{frontmatter}

\section*{Introduction}

\begin{figure*}[bt]
\psfig{file=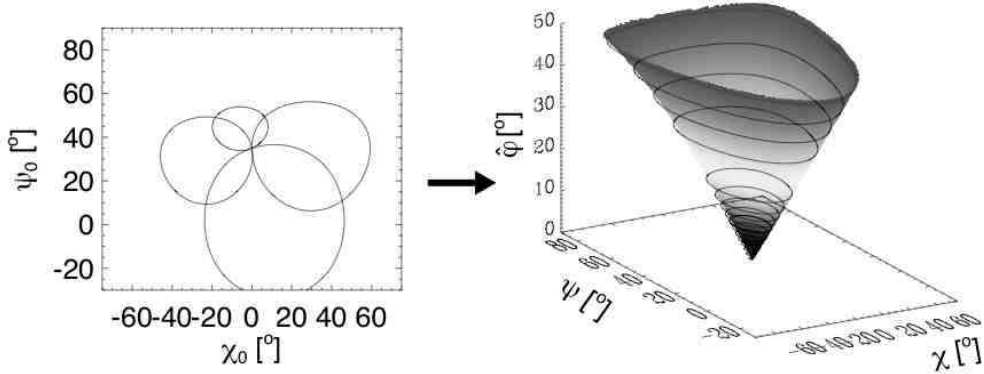,width=\linewidth,clip=}
\caption{ Event circles for a point source (left) and the
  corresponding response function in the 3D imaging data space of an
  ideal classical Compton telescope (right). A point source describes
  a cone with half-opening angle of 45\deg. The density along the
  scatter angle dimension corresponds to the Klein-Nishina cross
  section and describes the density of circles of radius $\varphi$ in
  the event circle representation \cite{UOberlack:97:phdthesis}.}   
\label{f:CT_dataspace}
\end{figure*}

The Liquid Xenon Gamma Ray Imaging Telescope (LXeGRIT) is a prototype
of Compton telescope (CT) based on a liquid xenon time projection
chamber (LXeTPC), which combines good energy resolution with imaging of
individual \MeV\ \g-ray interactions with submillimiter position
resolution (Sec.~\ref{sec:1}).     
A CT images \g-ray sources in the energy range from few 100~\keV\ to
more than 10 \MeV, reconstructing the direction of individual \g-rays
through Compton kinematics.      
The scatter angle on a free electron (\phibar) is given by the Compton 
formula  
\begin{equation}
\cos \phibar = 1 + \frac{1}{W_0} - \frac{1}{W_1} \textrm{,
  with:} 
  \quad W_i = \frac{E_i}{m_e c^2} 
\label{eq:compton-f}
\end{equation}
where $m_ec^2 = 0.511$~\MeV, $E_0$ is the initial energy and $E_1$ is
the energy of the scattered \g-ray. 
\begin{figure}[htb]
\centering
\psfig{file=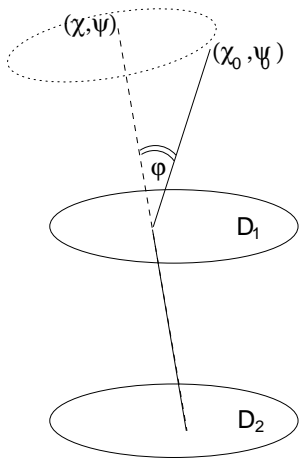,height=6cm,clip=}
\caption{Principle of a Compton telescope}
\label{f:CT_principle}
\end{figure}
Fig.~\ref{f:CT_principle} helps illustrating the principle of Compton
imaging. A \g-ray from a source at position ($\chi_0,\psi_0$) Compton
scatters in the plane D$_1$ by the (true) angle $\varphi$ and is stopped
in the plane D$_2$; $\varphi$ is estimated from
Eq.~\ref{eq:compton-f}. The time sequence of the interactions needs
also be known (Sec.~\ref{sec:2}).  
Interaction positions are measured in both planes and give the
direction of the scattered  photon ($\chi,\psi$) 
The direction of an individual \g-ray is then determined to an  
{\it event circle} with radius \phibar\ around the direction 
($\chi,\psi$). This ambiguity stems from the non-measurement of the 
direction of the scattered electron. 
After collection of many source events, intersection of all event
circles defines the source position (Fig.~\ref{f:CT_dataspace}). 
While back-projection of event circles on the sky is useful in
visualizing the basic measurement principle, it is not the optimum
method to perform Compton imaging. In fact, it doesn't
include the probability distribution for scatter angles
$d\sigma/d\phibar$ for photons from a source at given location and
energy, i.e. the frequency of occurrence of event circles with a given
opening angle. Fig.~\ref{f:CT_dataspace} illustrates how this
additional information can be exploited in a 3D data space consisting
of ($\chi, \psi, \phibar$), where a point source defines a cone-like
structure with half opening angle of 45\deg\ centered on the source
position, with a density along the \phibar\ dimension given by the
Klein-Nishina cross section convolved with factors that result from
the detector geometry and detector thresholds.    
Compton imaging of \g-sources is discussed in detail in
Sec.~\ref{sec:3} and examples are given in Sec.~\ref{sec:4}.  

\section{\label{sec:1} The LXeGRIT Compton Telescope}

For a detailed description of LXeGRIT and its performance as an 
imaging calorimeter, see
\cite{ACurioni:2004:PhDthesis,ACurioni:2006:nimeff,EAprile:98.nim}.   
In this section we summarize those aspects most relevant to Compton
imaging.        

\begin{figure}[htb]
\centering
\includegraphics[width=0.9\linewidth,clip]{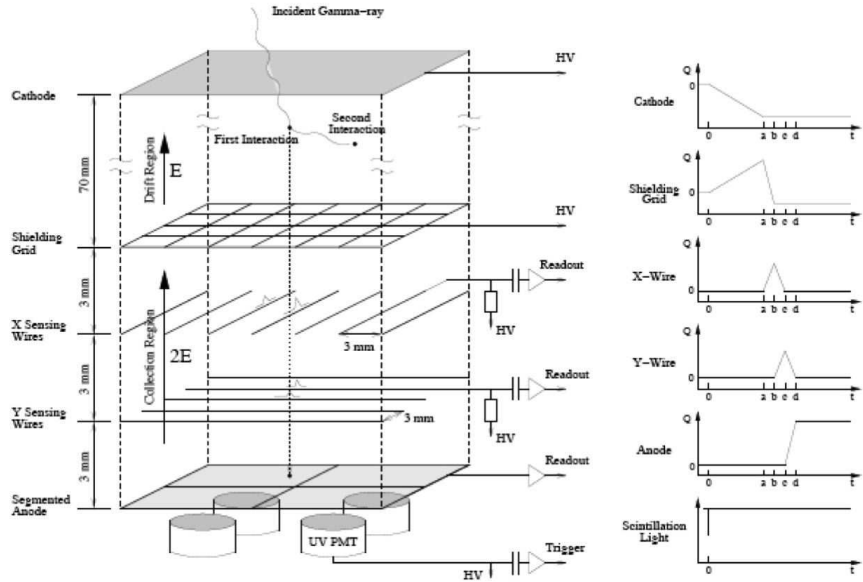}
\caption{Schematic of the LXeTPC readout structure (not to scale)
  with corresponding light trigger and charge pulse shapes. From
  \cite{ACurioni:2006:nimeff}. }   
\label{f:TPC_schematic}
\end{figure}


%
The fiducial volume of the LXeTPC is a box with dimensions
18.6$\times$18.6$\times$7~cm$^3$ (in the $x$, $y$, $z$ coordinates  
respectively), filled with high purity LXe.
At a temperature of $T=-95^{\circ}$~C the density of LXe is 2.85~\gcc
~\cite{NIST:chemistry:webbook:2005} and the attenuation length for
1~\MeV\ \g-rays in LXe is $\sim$6.2~cm ~\cite{NIST:XCOM:database}.
When a \g-ray interacts in the fiducial volume, both scintillation
light and ionization charge are produced efficiently, with W-values of 
$W_\mathrm{ph}=$ 24~eV \cite{doke:W_ph} and $W_\mathrm{e}$ = 15.6~eV 
\cite{takahashi:W_e}. 
The VUV (178~nm) scintillation photons are detected by four
photomultiplier tubes (PMTs), which provide the event trigger and the
initial time, t$_0$. The PMTs \footnote{2$''$ Electron Tubes
9813QA.} are coupled to the LXe volume via quartz windows.     
The ionization electrons drift under an applied electric field of
1~kV/cm, inducing a signal on two parallel wire planes, after passing
a Frisch grid. There are 62 wires in each plane and the pitch of the
wires is 3~mm. The location of the hit wire(s) in the two wire planes
provide the $x$ and $y$ coordinates in the TPC reference frame, while
the time, measured starting from t$_0$, gives the interaction depth
($z$ coordinate). The wires are transparent to the drifting charge,
which is finally collected by one of four independent anodes, and the
amplitude measures the energy deposited in the interaction. A
schematic of the readout structure and light trigger of LXeGRIT is
shown in Fig.~\ref{f:TPC_schematic}. 

A reliable Monte Carlo (MC) simulation of the detector has been
developed and tested to produce accurate results up to several
\MeV \cite{ACurioni:2006:nimeff}.  
For a generic \g-ray event, the LXeGRIT output is (E$_1$, x$_1$,
y$_1$, z$_1$); ...; (E$_n$,  x$_n$, y$_n$, z$_n$) where $n$ is the
event multiplicity, i.e. the number of detected interactions once the 
finite spatial resolution and energy threshold ($\sim$150~\keV\ for a 
single interaction) are accounted for.

\subsection{Position resolution}

\begin{figure}[htb]
\centering
\includegraphics[bbllx=55,bblly=505,bburx=550,bbury=730,
        width=.95\linewidth,clip]{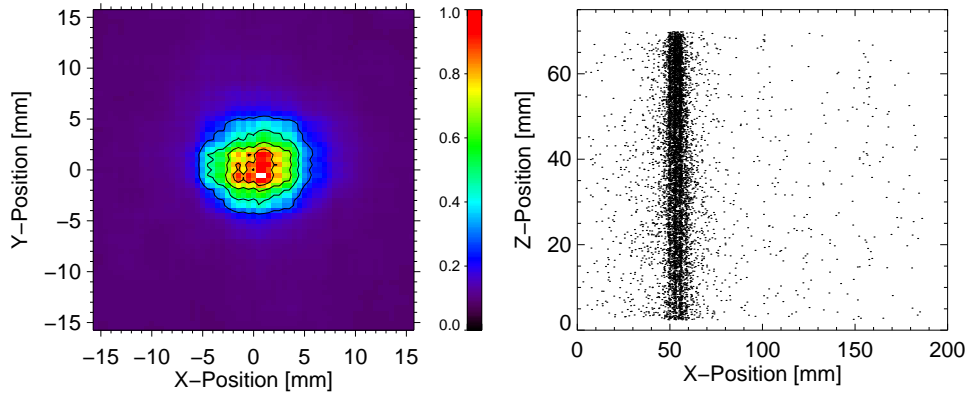}
\caption{Image of the \g-beam from a collimated \cs\ source on top of the
TPC.{\it Left:} projection in the $x-y$ plane. The coordinates have been
re-defined in order to have the source image centered at $x=0,~y=0$. {\it
Right:} side view ($x-z$ plane) of the same \g-beam. } 
\label{f:pore.2} 
\end{figure}

\begin{figure}[htb]
\includegraphics[bbllx=70,bblly=505,bburx=330,bbury=730,
        width=0.45\linewidth,clip]{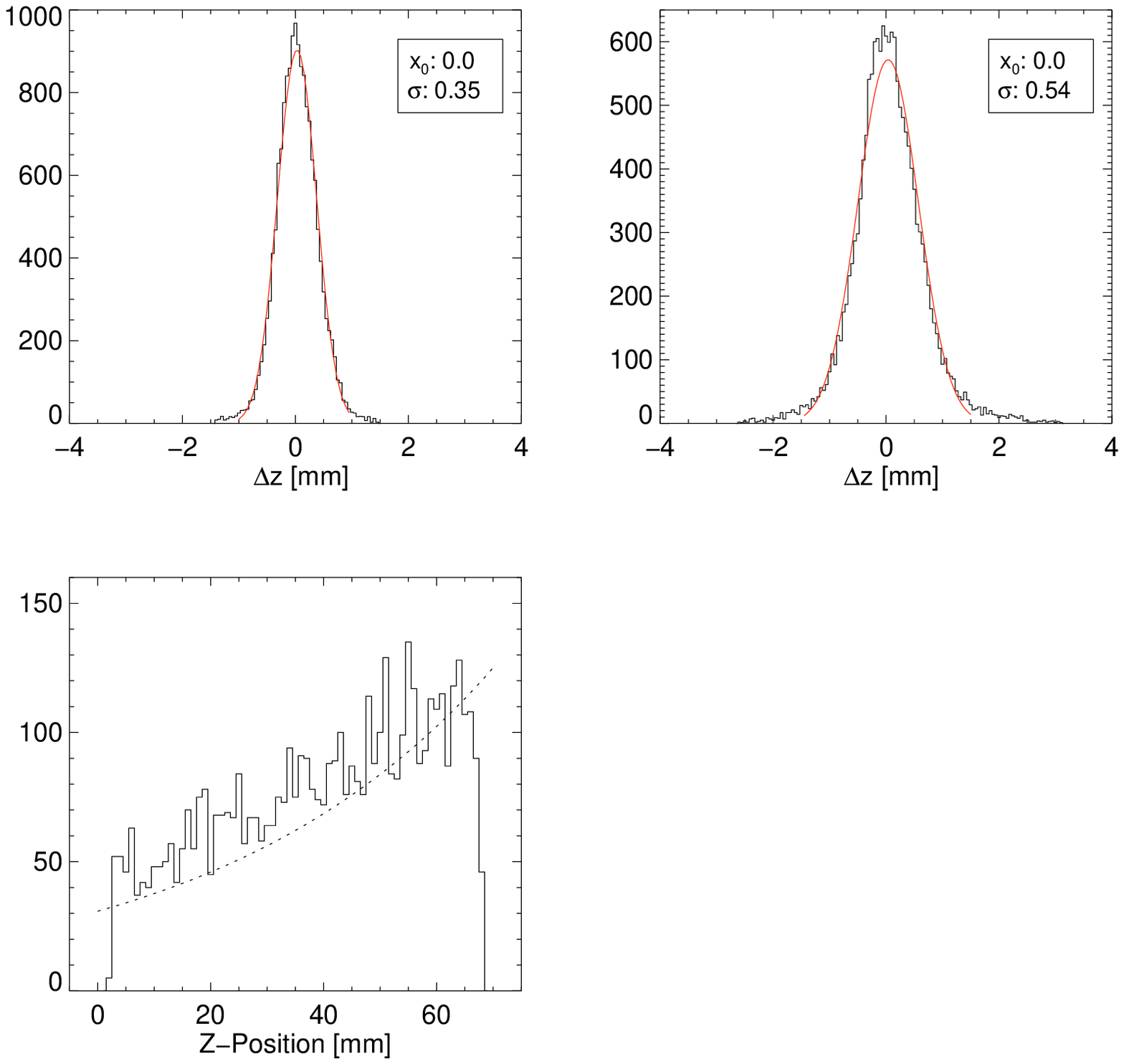}
\includegraphics[bbllx=70,bblly=280,bburx=330,bbury=505,
        width=0.45\linewidth,clip]{NIM_2003_8.ps}
\caption{{\it Left:} difference between the $z$-position as determined
  from the $x$ and $y$ wires. {\it Right:} $z$-distribution for a
  collimated \cs\ source sitting on top of the LXeTPC. Events in the
  full energy peak have been selected; superimposed ({\it dashed line})
  the exponential attenuation for 0.662~\MeV\ photons. The slight
  discrepancy at $z~\sim 60$~mm is  mainly due to the $z$ dependent
  light trigger efficiency, which is not corrected for.}    
\label{f:pore.1}
\end{figure}

The interaction location in the LXeTPC is obtained from the wire
signature. With a wire pitch of 3~mm, the spatial resolution is
0.87~mm ($\mathrm{rms} = 3~\mathrm{mm}/\sqrt{12}$) in $x-y$, if only a  
single wire signal is detected for each coordinate. For LXeGRIT this
is the typical case for energy deposits $<$0.3~MeV
~\cite{UOberlack:signal:recognition}. 
The reconstructed image of a collimated beam of \g-rays
(0.662~\MeV) photoabsorbed in the sensitive volume is shown in
Fig.~\ref{f:pore.2}, both in the $x-y$ and in the $x-z$ views. 
The \cs\ source was located above the TPC, collimated to a beam with 3
mm diameter in the $x-y$ plane by a lead collimator 15 cm thick. 
The position resolution in the $z$-coordinate is estimated to be about
0.25~mm (1 sigma). It is derived as the difference in drift times
measured independently on the $x$ and $y$ wires for the same
interaction, as shown in Fig.~\ref{f:pore.1}-$left$. Assuming two
independent measurement with Gaussian errors (neglecting the
uncertainty in the common trigger), $\sigma = 0.35 \mathrm{mm} /
\sqrt{2} \approx 0.25$~mm.   
The $z$-distribution of photoabsorbed events from the
same collimated \cs\ source is shown in Fig.~\ref{f:pore.1}-$right$,
well reproducing the expected exponential attenuation.
This spatial resolution fulfills the requirement of a fine grained CT,
where a typical separation between interactions is in the few cm range
and the linear extension of the charge cloud due to \MeV\ energy
deposits is typically less than 1~mm. 


\subsection{Energy resolution}

\begin{figure}[htb]
\centering
\includegraphics[bbllx=55,bblly=505,bburx=550,bbury=730,
        width=.95\linewidth,clip]{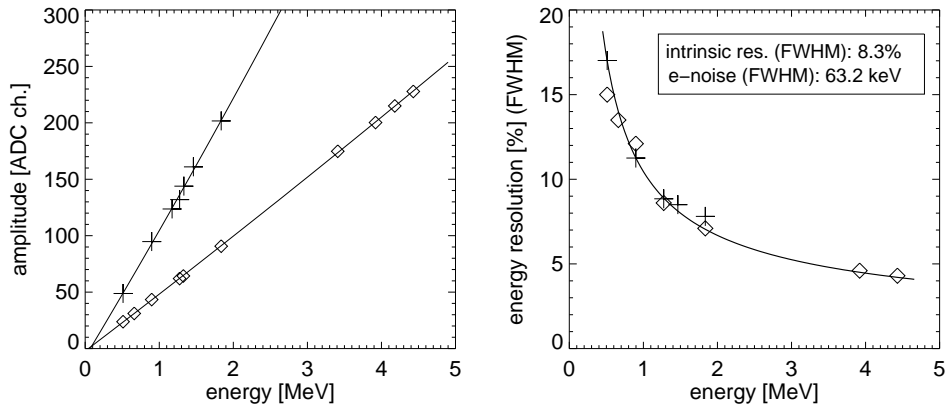}
\caption{{\it Left:} linearity plot for ADC channel vs. energy in
\MeV\ for both the 1999 ({\it open diamonds}) and the 2000 ({\it
crosses}) LXeGRIT electronics configuration. The gain in 2000 was
about twice the gain in 1999. {\it Right:} energy resolution 
versus energy, showing the $1/\sqrt{E}$ dependence expected from
Poisson statistic corrected by a constant term.}
\label{f:ecal.eres}
\end{figure}

The energy response is determined to be linear over the energy range
from 0.5 to 4.4 \MeV, covered by the available calibration sources
\footnote{\na (0.511 and 1.275 \MeV), \cs (0.662 \MeV), \yt (0.898 and
  1.836 \MeV), \co (1.173 and 1.332 \MeV), \pot (1.465 \MeV), Am-Be
  (4.43 \MeV).}, and is shown in Fig.~\ref{f:ecal.eres}-$left$. 
The energy dependence of the energy resolution is shown in
Fig.~\ref{f:ecal.eres}-$right$ and is described as
$$
\Delta E/E (FWHM)~=~\sqrt{\frac{P_1 ^2}{E} + \frac{P_2}{E^2}}
$$
where $P_1$ accounts for the intrinsic energy resolution from the
statistics of charge carriers in LXe; $1/\sqrt{E}$ reproduces the
energy dependence expected from Poisson statistic; $P_2$ accounts for
contributions which are independent of energy ({\it noise term}). 
The noise term parameterizes electronic noise,
errors in fitting the anode wave function, shielding inefficiencies of
the wires structure etc. In practice it is well described
considering the electronic noise only  (63~\keV\ FWHM,
which corresponds to about 1000 equivalent noise charge, as in
\cite{EAprile:98.nim}).    
The intrinsic energy resolution is 8.3\% at 1~\MeV, in good agreement
with previous measurements in LXe at the same drift field (1~kV/cm),
from gridded ionization chambers of much smaller fiducial volume
\cite{EAprile:nim91}. 

\section{\label{sec:2} Gamma ray tracking}

The most general interaction sequence useful for Compton imaging is a
first Compton scatter followed by full absorption of the scattered
\g-ray in one or more interactions
\footnote{For additional reading on \g-ray tracking, see e.g.
  \cite{SPIE:CSR,Boggs:Jean:2000,Schmid:1999,vanderMarel:Cederwall:1999}}.     
In LXeGRIT, a large fraction of multi-site events has only two
interactions (Compton scatter followed by photoabsorption). The
fraction of events with more than 3 interactions is at most 5\% of the
fully contained events for energies below 5~\MeV\ 
and is not considered in the following.   
The two cases for 2- or 3- interactions are treated separately in
Sec.~\ref{sec:2.1} and Sec.~\ref{sec:2.2}.

Events with interactions other than Compton scattering and
photoabsorption are not considered here for Compton imaging. The
relative fraction of these events increases with energy and becomes
dominant above $\sim$5~\MeV. 
The cross section for pair production turns on at 1.022~\MeV\ and, in
LXe, equals the Compton cross section at $\sim$6~\MeV.
Relativistic electrons lose energy both by ionization and by
radiation of secondary photons (Bremsstrahlung), which, if of
sufficiently high energy, may be detected at a 
separate location. For electrons the radiative energy loss overcomes
the one due to ionization above the {\it critical energy} E$_c =
610~\MeV ~/~ (Z+1.24) $ for liquids and solids \cite{PDG}. For Xe, $Z
= 54$ and E$_c = 11~\MeV$. Experimentally, we observe considerable
modifications to the energy spectrum for the 4.4~\MeV\ Am-Be
calibration source.      

\subsection{\label{sec:2.1} 2-site events}

\begin{figure}[htb]
\includegraphics[bbllx=90,bblly=505,bburx=310,bbury=730,
        width=0.30\linewidth,clip]{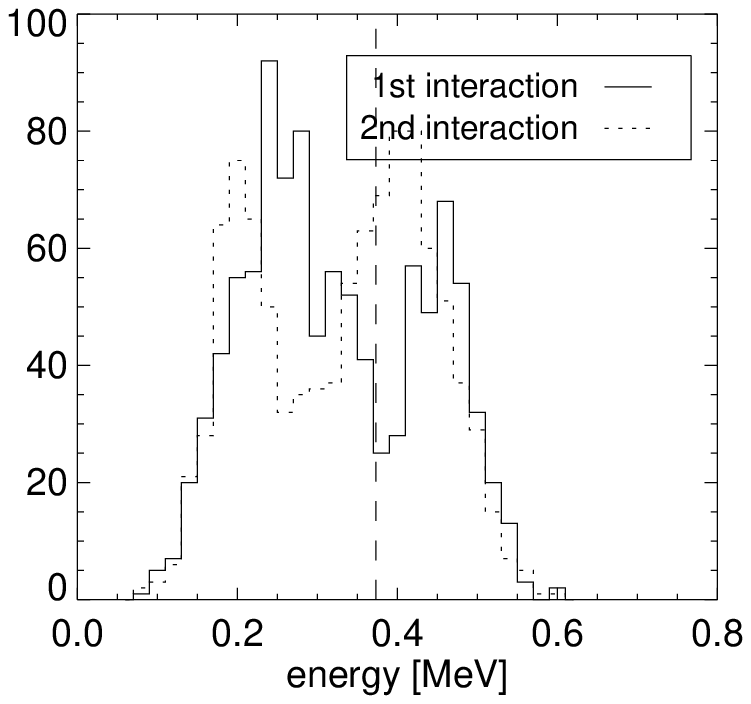}
\hfil
\includegraphics[bbllx=90,bblly=505,bburx=310,bbury=730,
        width=0.30\linewidth,clip]{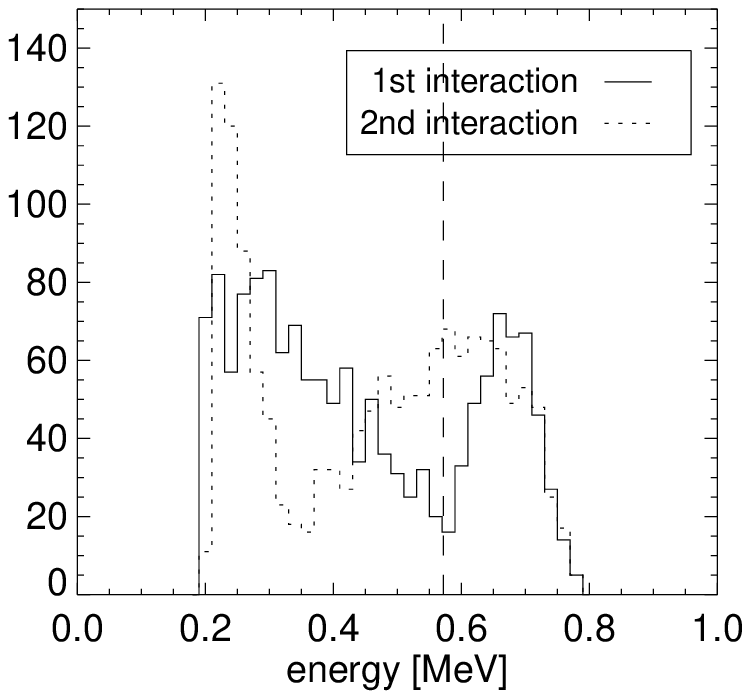}
\hfil
\includegraphics[bbllx=90,bblly=505,bburx=310,bbury=730,
        width=0.30\linewidth,clip]{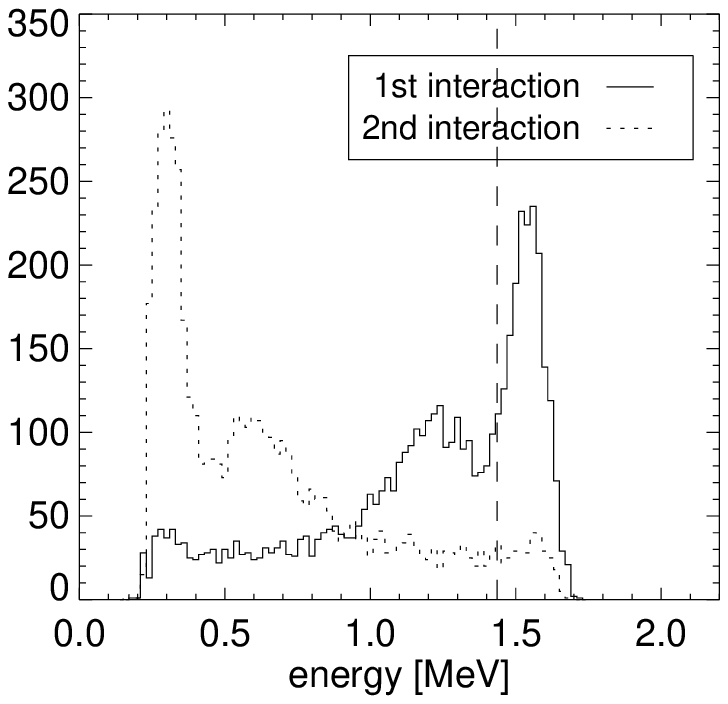}
\caption{ $E_1$, $E_2$ distribution for fully contained 2-site events. From 
left to right: 0.662~\MeV\ (\cs), 0.898 and 1.836~\MeV\ (\yt). }
\label{f:e1-e2:sources}
\end{figure}

\begin{figure}[htb]
\centering
\includegraphics[bbllx=55,bblly=505,bburx=570,bbury=730,
        width=.95\linewidth,clip]{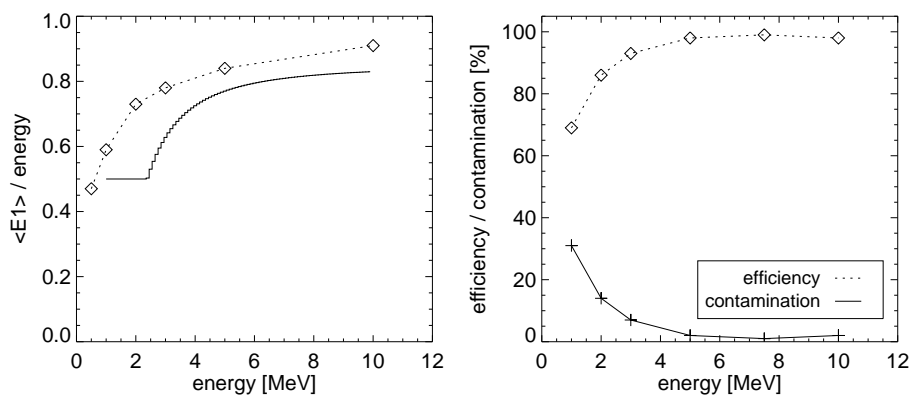}
\caption{  
{\it Left:} $ < E_1 > / E_{\gamma} $ vs. $E_{\gamma}$, from MC
data. Superimposed $f(E_{\gamma})$ as defined in
Eqs.~\ref{eq:fe.1}-\ref{eq:fe.2} 
{\it Right:} Efficiency and contamination in sequencing 2-site events
according to the procedure described in Sec.~\ref{sec:2.1}.
}
\label{f:e1-e2:1}
\end{figure}

In the case of 2-site events, assuming no prior knowledge the right
sequence can be guessed with a 50\% success rate. This improves using
the {\it energy sharing} between the two interactions that is, in many
cases, highly asymmetric.   
For LXeGRIT, the argument goes like this:
for energies larger than $\sim$2~\MeV\ a \g-ray is more likely to be
stopped in the fiducial volume with only two interactions if it does
lose a large fraction of the initial energy in the {\it first}
interaction. If the energy lost in the first interaction is small, the
scattered photon will most likely interact more than once before being
absorbed, i.e. would be classified as a 3(+)-site event.
Fig.~\ref{f:e1-e2:sources} shows the $E_1$ and $E_2$ energy spectra
for 0.662~\MeV\ (\cs), 0.898 and 1.836~\MeV\ (\yt) photons.  
The \cs\ source was collimated to a beam with a lateral spread of
$\sim$3~mm, which allows us to tag the first interaction as the
one within the collimator aperture.
The \yt\ source was at a distance of 2~m above the detector, without
any collimation. Since the source position is known, it is possible to
use Compton imaging (Sec.~\ref{sec:3}) and track each \g-ray assuming
the two possible sequences. The sequence which gives the correct
source position is then chosen as the true one. 
For 1.836~\MeV\ the $E_1$ and $E_2$ spectra almost mirror each
other (Fig.~\ref{f:e1-e2:sources}-$right$), and the situation $E_1 >
E_2$ is clearly the most likely. At lower energies (0.662 and
0.898~\MeV) the two spectra are much more similar and the $E_1 / E_2$
asymmetry is no more a good argument.   
The minimum in the $E_1$ spectrum, clearly visible for all the three
energies, corresponds to 90\deg\ scatter angle (the corresponding
$E_1~=~E^2_{tot}/(m_e c^2 + E_{tot}) $ is marked with a vertical
dashed line). 
It is a geometrical artifact, due to the direction of the incident
\g-rays along the detector $z$ axis, such that for a \g-ray scattered at
90\deg\ $z_1 \simeq z_2$, while a minimum separation of about 3~mm
along the $z$-axis is required to ensure a good energy determination
\cite{ACurioni:2004:PhDthesis}.    
In Fig.~\ref{f:e1-e2:1}-$left$ the ratio $ < E_1 > / E_{\gamma} $ is
plotted vs. $E_{\gamma}$ for MC data, where $< E_1 >$ is the mean of
the first energy deposition and $E_{\gamma}$ is the nominal energy of
the \g-ray. The trend is quite clear: 
$ < E_1 > / E_{\gamma} $ increases with $E_{\gamma}$ and $ < E_1 > /
E_{\gamma} ~\geq$~75\% for $E_{\gamma} ~\geq$~2~\MeV.
Fig.~\ref{f:e1-e2:1}-$left$ is suggestive of an optimized selection on
the energy sharing, requiring the ratio $E_1 / E_{tot}$ to be larger
than some value $f(E_{tot})$
\begin{equation}
f(E_{tot}) = 0.85 \left( 1 - \frac{2}{E_{tot} ^2} \right)~~;~E_{tot} >
2.4~\MeV 
\label{eq:fe.1}
\end{equation}
\begin{equation}
f(E_{tot}) = 0.5 ~~;~1~\MeV \leq E_{tot} \leq 2.4~\MeV 
\label{eq:fe.2}
\end{equation}
shown in Fig.~\ref{f:e1-e2:1}-$left$ with $E_{tot} = E_{\gamma}$.  
Fig.~\ref{f:e1-e2:1}-$right$ gives efficiency and contamination for
this sequencing procedure in the energy range 1-10~\MeV. The
efficiency is as high as 86\% at 2~\MeV\ and saturates ($\geq$98\%) at 
5~\MeV.  
Below 1~\MeV\ no optimum criterion for sequencing 2-site events has
been found. 

\subsection{\label{sec:2.2} Multi-site events } 

When more than two interactions are available, the time sequence is,
in principle, univocally determined by Compton kinematics. 
In the general case of a \g-ray which undergoes $N-1$ Compton scatters
and is photoabsorbed in the $N^{th}$ interaction energy and momentum
conservation is written as   
\begin{equation}
  E~^{\gamma}_{i-1} = E~^{\gamma}_{i} + E~^\mathrm{e}_i ~~~~~~;~~~
  \vec{p}~^{\gamma}_{i-1} = \vec{p}~^{\gamma}_{i} + \vec{p}~^\mathrm{e}_i  
  \label{e:momentum}
\end{equation}
with $E_{i}^{\gamma} \; (i=0 ,\ldots, N-1)$ and $E_i^\mathrm{e}\; (i=1
,\ldots, N)$ the energy of the \g-ray and the scattered electron after
interaction $i$; $E_{0}^{\gamma}$ is the energy of the incoming photon
and $\vec{p}_{i}$ are the corresponding momenta.
The electron scatter angle is not measured and is ignored in the
following. For the photon scatter angle $\varphi$
\begin{equation}
  \cos \varphi_i  = 1 + \frac{1}{W_i} - \frac{1}{W_{i+1}} \textrm{,
  with:} 
  \quad W_i = \frac{E~^{\gamma}_i}{m_0 c^2} 
  \label{e:phi} 
\end{equation}
For a given interaction sequence, the interaction locations determine
geometrically $N-2$ photon scatter angles \phigeo$~_i \ (i=2,\ldots,
N-1)$: 
\be
\cos{\varphi_{\mathrm{geo} i}} = \frac{\overrightarrow{u} _i \cdot
\overrightarrow{u} _{i+1}}{|\overrightarrow{u} _i | 
|\overrightarrow{u}_{i+1} |}   
\ee 
where $\overrightarrow{u} _i ~=~ (x_i-x_{i-1},\: y_i-y_{i-1},
\:z_i-z_{i-1})$.  

$N-1$ Compton scatter angles \phibar$_i$ are measured by the
energy deposits according to equation~\ref{e:phi}, noting that
$E_i^{\gamma} = \sum_{j=i+1}^{N} E_j \; (i = 0,\ldots,N-1)$. This
redundant information allows testing of the sequence of the
interaction points based solely on kinematics. A straightforward test
statistic consists in summing the differences of the scatter angles
quadratically, weighting the summands with the measurement errors:  
\begin{eqnarray} \label{e:testa}
  T_\varphi &=& \frac{1}{N-2} \sum_{i=2}^{N-1} \frac{(\cos{\phibar_i}
  - \cos{\varphi_{\mathrm{geo}~i}})^2}{\sigma_i^2} \\ 
  \textrm{with:} \quad \sigma_i^2 &=& \sigma_{\cos{\phibar},i}^2 +
  \sigma_{\cos{\phigeo},i}^2 
  \nonumber 
\end{eqnarray}
Ideally, the test statistic would be zero for the correct sequence if
the photon is fully contained. With measurement errors,
$T_\varphi^\prime$ is always greater than zero, but the correct
interaction sequence is still most likely to produce the minimum value
of the test statistic. A straightforward interpretation as a reduced
$\chi ^2$ distribution is not possible due to the non-Gaussian
shape of the probability distribution in \phibar. 
For each triplet of interactions \ephibar \ and \ephigeo \ then are
computed 
 
\begin{eqnarray} \label{e:errors}
\sigma_{\cos{\phigeo},i}^2 &=&
\sum_{k=1}^{3} \Bigg\{ \left ( \frac{u_{i+1,k}}{|\overrightarrow{u}
  _i| \cdot |\overrightarrow{u}_{i+1}| } -
\frac{u_{i,k}~\cos{\phigeo}}{|\overrightarrow{u} _i| ^2} \right )^2 +  
\nonumber \\ 
&& 
\left ( \frac{u_{i,k}}{ |\overrightarrow{u} _i| \cdot
|\overrightarrow{u}_{i+1}| } - 
\frac{u_{i+1,k}~\cos{\phigeo}}{|\overrightarrow{u} _{i+1}| ^2} \right
)^2 
\Bigg\} 
\cdot \sigma _k ^2 \\ 
\textrm{with:} &k& \ \textrm{spatial coordinate index and} \nonumber \\ 
&\sigma _k & \ \textrm{position uncertainty on each coordinate}
\nonumber \\  
\sigma_{\cos{\phibar},i}^2 &=& 
\frac{1}{W_i ^4}\cdot \sigma (W_i - W_{i+1}) ^2 + \left
( \frac{1}{W_i ^2} - \frac{1}{W_{i+1} ^2} \right) ^2 \cdot \sigma
(W_{i+1}) ^2 
\end{eqnarray}

\begin{figure}[htb]
\centering
\includegraphics[bbllx=55,bblly=505,bburx=570,bbury=730,
        width=\linewidth,clip]{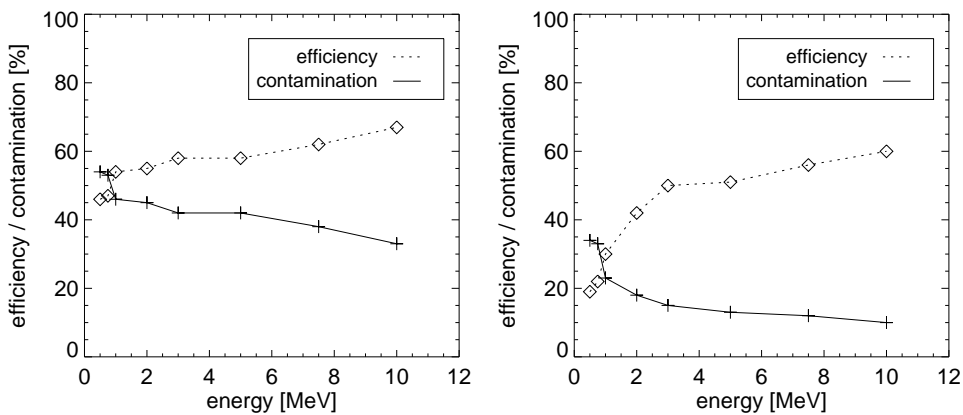}
\caption{ Efficiency and contamination in the reconstruction of the
correct time sequence for 3-site event, as calculated using MC data
and the actual algorithm used for experimental data.   
{\it Left:} without applying any further selection. In this case events can only
be {\it correctly} or {\it wrongly} sequenced, since no event is rejected. 
{\it Right:} selecting $E_1 / E_{tot} > 0.3$ {\it a posteriori}. The
contamination fraction is here defined as the fraction of events wrongly
sequenced which have $E_1 / E_{tot} > 0.3$. 
}
\label{f:CSR-eff:2}
\end{figure}

We consider here only the case of 3 interactions, by far
the most likely. There are six (3!) possible sequences to start with, 
i.e. assuming no additional knowledge the right sequence is chosen
17\% of the times.
The efficiency of this procedure is shown in
Fig.~\ref{f:CSR-eff:2}-$left$, using MC data.  
It is $\sim$55\% at 2~\MeV\ and exceeds 60\% above 5~\MeV. In LXeGRIT
the energy resolution is the limiting factor. 
The fraction of wrongly sequenced events ({\it contamination}) is also
shown in Fig.~\ref{f:CSR-eff:2}-$left$. In this case, since no event
is rejected, it is just the complement to 1 of the
efficiency. 
It is worthwhile to note that the most frequent confusion of the
interaction sequence involves the swap of second and third
interaction, while the first interaction is properly found. Such
events, counted here under ``confusion'', remain usable for imaging if  
the separation between second and third interaction is considerably
shorter than the separation of first and second interaction. In this
case, the wrong sequence leads to tails in angular resolution but not
to a ``conversion'' of source photons into background photons.  
Assuming the efficiency shown in Fig.~\ref{f:CSR-eff:2}-$left$ as an
upper limit, the algorithm performance can be improved applying
further selections {\it a posteriori}, with the goal of keeping the 
efficiency as close as possible to the one in
Fig.~\ref{f:CSR-eff:2}-$left$ while reducing the contamination.    

\begin{figure}[htb]
\centering
\includegraphics[bbllx=55,bblly=505,bburx=570,bbury=730,
        width=\linewidth,clip]{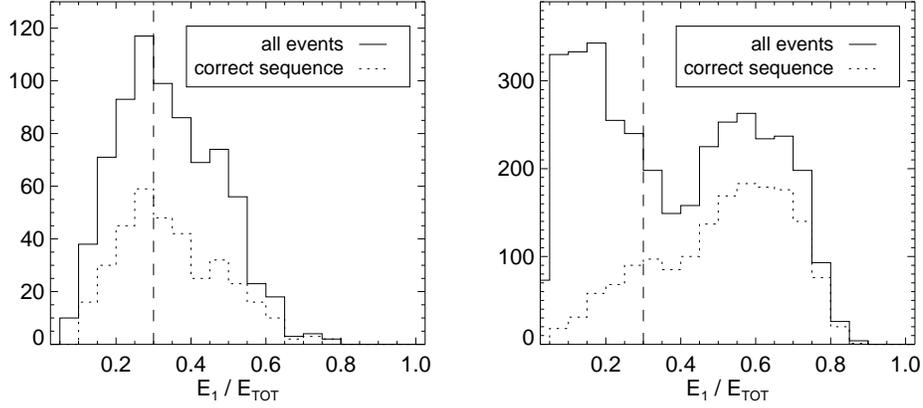}
\caption{ $E_1 / E_{tot}$ for 0.898 ({\it left}) and 1.836 ({\it
    right}) \MeV\ \g-rays. From experimental data, imposing full
    energy containment. } 
\label{f:CSR-eff:3}
\end{figure}

The most powerful variable is $E_1 / E_{tot}$, where $E_1$ is 
the energy deposited in the {\it first} interaction \footnote{{\it
    First} according to the \g-tracking algorithm}.
Fig.~\ref{f:CSR-eff:3} shows $E_1 / E_{tot}$ for the 0.898 and
1.836~\MeV\ \yt\ line (experimental data), when all the events are
considered and selecting only events with the time sequence correctly
identified, i.e. events for which the source is correctly imaged.
At 0.898~\MeV\ there is no clear correlation between $E_1 / E_{tot}$
and finding the right sequence, but at 1.836~\MeV\ a large fraction of
wrongly reconstructed events shows up at $E_1 / E_{tot} < 0.3$. The
impact of selecting $E_1 / E_{tot} > 0.3$ has been studied over the
energy range 0.5-10~\MeV\ using MC data. 
The efficiency of the \g-tracking procedure combined with this {\it a
posteriori} selection is shown in Fig.~\ref{f:CSR-eff:2}-$right$
together with the contamination, here defined as the fraction of
events (in the full energy peak) which are wrongly sequenced and have
$E_1 / E_{tot} > 0.3$. 
This technique works well above 2~\MeV, while at lower energies the
reduction in efficiency combined with poor rejection power makes it
counterproductive, as also seen in Fig.~\ref{f:CSR-eff:3}-$left$.


\section{\label{sec:3} Compton imaging}

\subsection{\label{sec:3.1} Angular resolution}

If the source position is known, two independent measurements of the
first Compton scatter angle (\phibar\ and \phigeo) are given. The
difference between the two gives the measure of the angular resolution
(ARM). In this section ARM and 1-$\sigma$ angular resolution are
interchangeable.   
Three sources of error limit the angular resolution: {\it energy
resolution}, which limits the precision in measuring \phibar;
{\it position resolution}, which limits the precision in measuring
\phigeo; and {\it Doppler broadening}.  

\begin{figure}
\centering
\includegraphics[bbllx=55,bblly=505,bburx=570,bbury=730,
	width=\linewidth,clip]{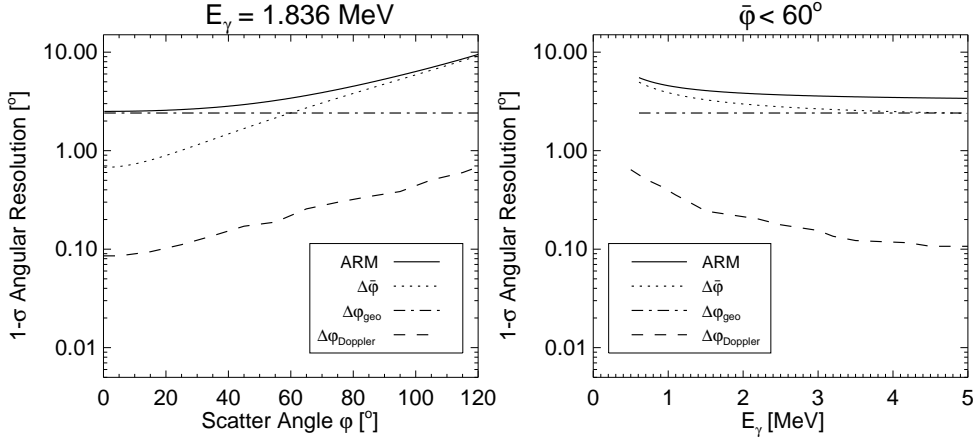}
\caption{ 
Expected angular resolution for LXeGRIT; $\Delta \phibar$ and $\Delta
\phigeo$ have been obtained using
Eqs.~\ref{eq:d-phibar},~\ref{eq:d-phigeo1} and combined to give the
final ARM according to Eq.~\ref{eq:d-phi}.  
{\it Left:} expected angular resolution vs. scatter angle $\varphi$
for a fixed energy (1.836~\MeV). 
{\it Right:} expected angular resolution for LXeGRIT vs. energy,
selecting \phibar\ $<$60$^{\circ}$, i.e. forward scattering. 
}
\label{f:ang_res}
\end{figure}

The uncertainty on the scatter angle due to the energy resolution is: 
\begin{equation}
\Delta\phibar = \frac{m_ec^2}{\sin\phibar}\sqrt{\left( \frac{\Delta
E_1}{E^2_{tot}} \right)^2 + \left( \frac{E_1(E_1 + 2E_2) \Delta
    E_2}{E^2_{tot} E^2_2} \right)^2 }  
\label{eq:d-phibar}
\end{equation}
where $E_{tot}$ is the initial energy of the \g-ray, $E_1$ the energy
deposited in the first interaction, $E_2 = E_{tot} - E_1$; in
Eq.~\ref{eq:d-phibar} there are only two free parameters,
e.g. \phibar\ and $E_{tot}$. Its behavior as a function of \phibar\
and of $E_{tot}$ is shown in Fig.~\ref{f:ang_res}. The curve for the
\phibar\ dependence is obtained for $E_{tot}$=1.836~\MeV, the one for
the $E_{tot}$ dependence is obtained integrating over \phibar\ $<$
60\deg\ according to the Klein-Nishina cross section. An energy
threshold of 150~\keV\ and an energy resolution of 10\% /
$\sqrt{E~[\MeV]}$ (FWHM) have been assumed.  

The uncertainty on the direction of the scattered \g-ray,
$\Delta\phigeo$, assuming a separation between the first two
interaction locations $|\overrightarrow{u}| $ large compared to
$\sigma _x $, $\sigma _y $ and $\sigma _z $, is given by   
\begin{equation}
\Delta\phigeo  =  \frac{ \sqrt{2} }{|\overrightarrow{u}| } \sqrt{
( \Delta\phigeo ) _x ^2 +
( \Delta\phigeo ) _y ^2 +
( \Delta\phigeo ) _z ^2 }
\label{eq:d-phigeo1}
\end{equation}
with $( \Delta\phigeo ) _a  = \frac{ \sqrt{2} }{|\overrightarrow{u}| }
\sigma _a \sqrt{ 1 - \left( \frac{\overrightarrow{u} \cdot \hat{a}
}{|\overrightarrow{u}| } \right) ^2 }, a=x,y,z.$ 
In Fig.~\ref{f:ang_res} $\Delta\phigeo$ is shown for
$|\overrightarrow{u}| = 30$~mm. 

The Compton formula in Eq.~\ref{eq:compton-f} gives the scattering
angle {\it if} the incident photons were to interact with stationary
free electrons. 
Doppler broadening constitutes an irreducible limitation to angular
resolution for a CT and its effect is larger for target materials of
larger atomic number, such as Ge or Xe, compared to Si or liquid
scintillators \cite{zoglauer:2003}. For low energy (few 100~\keV)
\g-rays, the uncertainty in the scatter angle due to Doppler
broadening contributes significantly to the overall ARM.  
Once energy  resolution and position resolution are taken into
account, Doppler broadening plays a rather negligible role for LXeGRIT
(Fig.~\ref{f:ang_res}), which is designed to image \g-rays of energy
0.5~\MeV\ or higher.

\begin{figure}[htb]
\centering
\includegraphics[bbllx=55,bblly=505,bburx=570,bbury=730,
	width=\linewidth,clip]{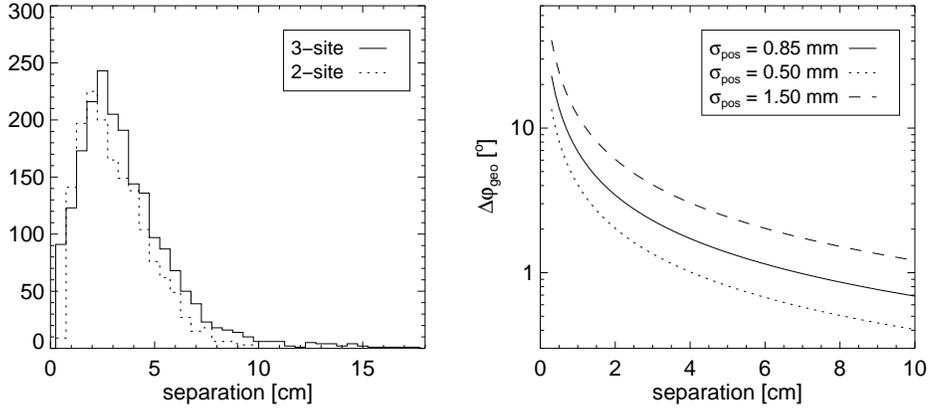}
\caption{ 
{\it Left:} 3D separation between the first and the second interaction
for 1.836~\MeV\ fully contained \g-rays.
{\it Right:} Angular spread $\Delta \phigeo$ vs. 3D separation for
different values of the position resolution. LXeGRIT achieves a
position resolution of 0.85~mm.}
\label{f:ARM:sepa}
\end{figure}

Neglecting Doppler broadening, the overall angular resolution $\Delta
\varphi$ is defined as  
\begin{equation}
\Delta\varphi = \sqrt{\Delta\phigeo ^2 + \Delta\phibar ^2}
\label{eq:d-phi}
\end{equation} 
At 1.836~\MeV\ the 1~$\sigma$ angular resolution is about 3$^{\circ}$
for scatter angles up to 60$^{\circ}$, improving for more forward
scattering.  
The dependence of $\Delta \varphi$ on the interaction separation is
shown in Fig.~\ref{f:ARM:sepa}-$right$, for different position
resolution. Given the typical separation of the order of few cm shown
in Fig.~\ref{f:ARM:sepa}-$left$, a mm position resolution is required
for a good imaging performance. 


\begin{figure}[htb]
\centering
\includegraphics[bbllx=55,bblly=505,bburx=570,bbury=730,
	width=\linewidth,clip]{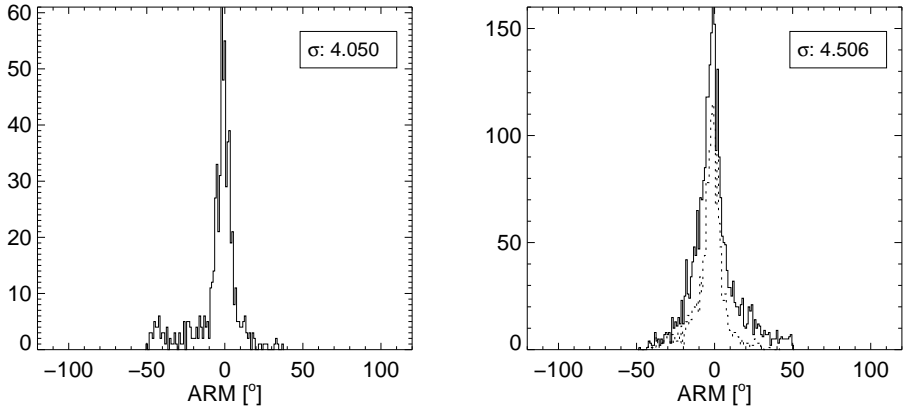}
\caption{ 
ARM spectra for 1.836~\MeV\ \g-rays (\yt\ source). 
The standard deviation has been obtained fitting the ARM spectra with
a Gaussian function.  
{\it Left:} 3-site events.
{\it Right:} 2-site events. The dashed line indicates events with \phibar\
restricted to less than 70\deg. This selection makes the ARM distribution
sensibly narrower, getting rid of the extended tails due to large scatter
angles. The standard deviation has been derived from this selected sample.
}
\label{f:ARM:arm}
\end{figure}

\begin{figure}[htb]
\centering
\includegraphics[bbllx=55,bblly=505,bburx=570,bbury=730,
	width=\linewidth,clip]{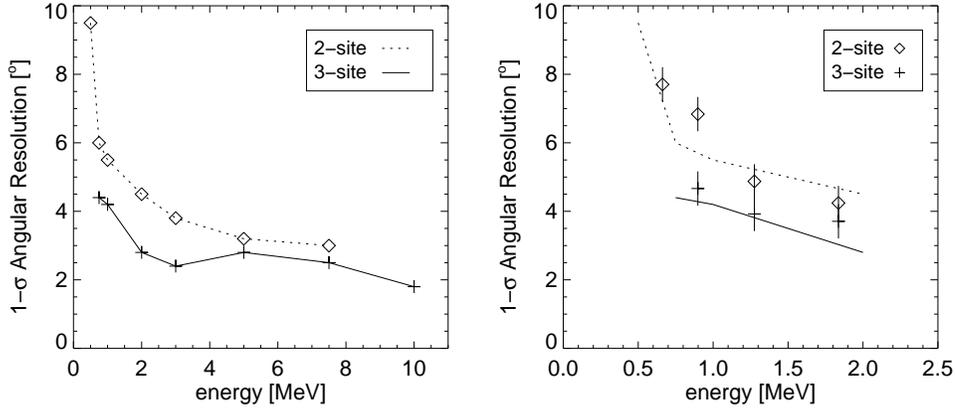}
\caption{ 
ARM spread vs. energy obtained using accurate MC data and experimental
data. The 2- and 3-site samples are shown separately. 
{\it Left:} MC data; the point at 0.5~\MeV\ in the 3-site data and the
one at 10~\MeV\ in the 2-site data have been omitted because of very
little statistical significance. Full energy containment has been
imposed.  
{\it Right:} Experimental data: 0.662 (\cs), 0.898 (\yt), 1.275 (\na)
and 1.836 (\yt) \MeV\ for 2-site events, 0.898, 1.275 and 1.836~\MeV\
for 3-site events. The corresponding MC curves have been superimposed.
}
\label{f:ARM:1}
\end{figure}

The ARM spectra for real data, 1.836~\MeV\ \g-rays, is shown in
Fig.~\ref{f:ARM:arm}. For a realistic comparison with expectation, 
the analysis is based on MC data. In this way it is also possible to
separate the response for 2- and 3-site events. 
The result for the energy band 0.5-10~\MeV\ is shown in
Fig.~\ref{f:ARM:1}-$left$. In Fig.~\ref{f:ARM:1}-$right$ MC data and
experimental data are compared for the lines: 0.662 (\cs), 0.898
(\yt), 1.275 (\na) and 1.836 (\yt) \MeV\  (2-site events), and 0.898,
1.275 and 1.836~\MeV\ (3-site events). 


\begin{figure}[htb]
\centering
\includegraphics[bbllx=55,bblly=505,bburx=350,bbury=730,
        width=0.55\linewidth,clip]{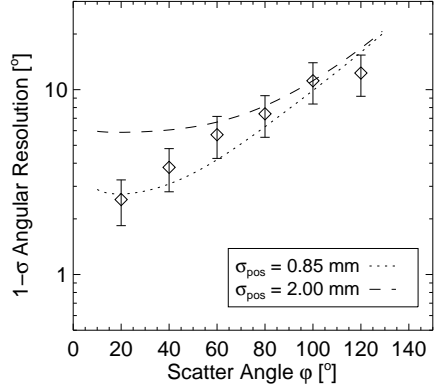}  
\caption{ Angular resolution vs. scatter angle for a sample of
  1.836~\MeV\ \g-rays, in \phibar\ bins of $20^\circ$. Superimposed,
  the expected angular resolution assuming two different values for
  the position resolution. The expected  LXeGRIT position resolution
  of about 0.85~mm agrees well with the data. A 2~mm position
  resolution, while still in good agreement with the data for scatter
  angles larger than 60\deg, is clearly ruled out by the two points at
  20\deg and 40\deg, the only ones actually sensitive to $\Delta
  \phigeo$. }  
\label{f:ARM:3}
\end{figure}

The angular resolution vs. \phibar\ is shown in Fig.~\ref{f:ARM:3},
from the same data (2- and 3-site events combined); the experimental
points are compared to the expected angular resolution (as shown in
Fig.~\ref{f:ang_res}-$left$), showing a good agreement. The
expectation for a position resolution degraded to 2~mm is also shown;
for small \phibar\ the overall performance is compromised.  

\subsection{\label{sec:3.2} Maximum Likelihood Image Reconstruction}

The imaging problem deals with the derivation of the intensity
distribution of the {\it object region} from the observational data,
which, for a Compton telescope, can be represented in a 3D binned data
space consisting of the scatter direction ($\chi$,$\psi$) and the
scatter angle \phibar: 
\begin{equation}
D_i = \sum_{j} R_{ij} f_j + b_i + N_i
\label{eq:imaging-1}
\end{equation}
where $D_i$ is the observed number of counts, $b_i$ the expected
background, and $N_i$ the statistical noise in the $i^{th}$ bin in
data space, $R_{ij}$ is the instrumental response, i.e. the
probability to detect a photon from the $j^{th}$ pixel 
in the object region in the $i^{th}$ bin in the data space, and $f_j$ is
the flux in the $j^{th}$ pixel in the object region
We restrict ourselves in the following to maximum likelihood fitting
of single or few point sources (plus background), scanning and testing
a grid of image pixels for one source at a time. 
The probability of the observed data under a specific model $\{f_j,
b_i\}$ is given by the likelihood function $L$, defined by multiplying
the probability of each bin   
\begin{equation}
L = \prod _i P_i  = \prod _i P(D_i | {f_j}, b_i)
\label{eq:imaging-3}
\end{equation}
As a counting experiment with fixed observing time the statistics in
each bin is given by the Poisson distribution 
\begin{eqnarray}
P_i & = & \frac{\omega _i ^{D_i}}{D_i !} ~e^{-\omega
_i}  ~~~
\textrm{for}~\omega_i > 0 \nonumber \\
P_i & = & 1 ~~~~~~~~~~~~~ \textrm{for}~\omega_i = 0,~ D_i = 0 
\nonumber \\ 
P_i & = & 0 ~~~~~~~~~~~~~ \textrm{for}~\omega_i = 0,~  D_i > 0
\label{eq:imaging-2}
\end{eqnarray}
where $\omega _i  = \sum _j R _{ij} f_j + b_i $ is the expected number
of counts in the $i^{th}$ bin and $P_i$ the probability of having
$D_i$ counts in the $i^{th}$ bin, given $\omega _i$.

Taking the logarithm of the likelihood function, one obtains
\begin{equation}
\log{L} = \sum _i D_i \log{\omega _i} - \sum _i \omega _i + C
\label{eq:imaging-4}
\end{equation}
where $C = - \sum _i \log{(D_i !)}$ is a constant with respect to the 
parameters $f_j$ and is therefore model independent and can be
neglected.    
Maximizing $L$ with respect to the flux distribution $\{f_j\}$
for the intensity $f_j$ results in the following set of  equations: 
\begin{equation}
\sum _i D_i \frac{\partial \omega _i}{\partial f_j} / \omega _i - \sum _i
\frac{\partial \omega _i}{\partial f_j} = 0
\label{eq:imaging-5}
\end{equation}
This is the general expression for maximum likelihood in binned mode.
A possible choice to solve the system in Eq.~\ref{eq:imaging-5} is the
Newton-Raphson algorithm \cite{WHPress:92}, which separately estimates
each pixel on the source parameters. The statistical significance is
then obtained from $-2\log{\lambda}$, where $\lambda$ is
the likelihood ratio of the two hypotheses background only and
background plus source \cite{HdeBoer:92}. $-2\log{\lambda}$ follows a
$\chi^2 _3$ distribution for an unknown point source and  a $\chi^2
_1$ distribution for a known source. 

A different approach, known as {\it list mode} likelihood method, aims
at reconstructing an image on an event-by-event basis. This method
may be derived from the binned likelihood method by increasing the
number of bins in the data space until each bin contains either 0 or
1 event. If only bins that contain an event are considered,
Eq.~\ref{eq:imaging-5} deals with events rather than bins.  
The probability to detect a photon from the $j^{th}$ pixel of
the object region in the $i^{th}$ bin turns into the probability
density for the $i^{th}$ photon in the data space. The list mode
maximum likelihood method is very useful in overcoming problems of
storage in computer memory and in reducing the CPU time needed for the
calculation for cases of sparsely populated dataspaces, e.g., when
each event contains multiple parameters relevant to the imaging
problem. A detailed description of this technique is given in
\cite{HHBarrett:97}.   

\begin{figure}[htb]
\centering
\includegraphics[bbllx=55,bblly=505,bburx=550,bbury=730,
        width=0.95\linewidth,clip]{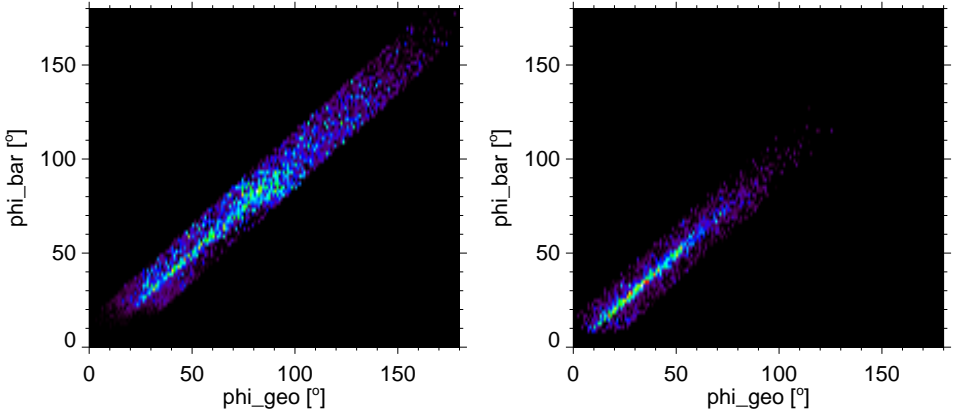}
\caption{$R^{(2)} (\phibar | \phigeo)$ or point-spread function (PSF)
for LXeGRIT, as obtained for MC data for a Crab-like source (i.e. with
a power law energy spectrum with index 2 in the energy band from 1 to
10 \MeV) 25\deg\ off-axis. The PSF is given  by the probability of
detecting \phibar\ (phi\_bar) for each \phigeo (phi\_geo). {\it Left:}
2-site events. {\it Right:} 3-site events. } 
\label{f:PSF}
\end{figure}

In the case of a CT, the equation in list mode can be derived
directly from the one in binned mode. One defines the source position
($\chi _0, \psi _0$) in a 3D data space (\phibar, $\chi$, $\psi$),
where $\chi$, $\psi$ is some reference frame, e.g. longitude and
latitude or right ascension and declination (Ra, Dec). The instrument
response is equivalently given in a 3D data space as $R^{(3)}(\chi,
\psi, \phibar)$  or in a 2D data space as $R^{(2)}(\phibar, \phigeo)$.  
In the 3D data space $R^{(3)}(\chi, \psi, \phibar | \chi _0, \psi _0)$   
has a conical shape with a half-opening angle of 45\deg and the vertex 
at the source location ($\chi _0$,
$\psi_0$) (Fig.~\ref{f:CT_dataspace}).
$R^{(2)}(\phibar|\phigeo)$ is given by the probability of measuring
\phibar\ for a given scatter angle \phigeo; an example for LXeGRIT is
shown in Fig.~\ref{f:PSF}. The probability distribution is enhanced
along the diagonal, i.e. for \phibar~=~\phigeo, which is equivalent to
having the ARM peak at 0\deg.   
$R^{(3)}$ and $R^{(2)}$ are connected through the relation
\begin{equation}
R^{(3)}(\chi, \psi, \phibar | \chi _0, \psi _0)~2\pi
\sin{\phigeo}d\phigeo~=~R^{(2)}(\phibar | \phigeo)~\cos{\psi}d\chi
d\psi  
\label{eq:imaging-6}
\end{equation}
We have presently implemented the list-mode likelihood method only
without the background term, i.e. the expected number of counts in the  
$i^{th}$ bin is now  
\begin{equation}
\omega _i = \sum _j R^{(3)} _{ij} f_j 
\label{eq:imaging-7}
\end{equation}
The logarithm of the likelihood function is written as  
\begin{equation}
\log{L} = \sum _i D_i \log{( \sum _j R^{(3)} _{ij} f_j )} - \sum _{i,j}
R^{(3)} _{ij} f_j + C
\label{eq:imaging-8}
\end{equation}
If the bin size is reduced until each bin has at most one count, $D_i$
can only be 0 or 1. The first term in Eq.~\ref{eq:imaging-8} is now
the sum over events rather than bins. Introducing an index $ie$ to
indicate events and $id$ to indicate bins, Eq.~\ref{eq:imaging-8} can
be rewritten as  
\begin{equation}
\log{L} = \sum _{ie} \log{(\sum _j R^{(3)} _{ie,j} f_j )} - \sum
_{id,j} R^{(3)} _{id,j} f_j + C 
\label{eq:imaging-9}
\end{equation}
Approximating $R^{(2)}(\phibar, \phigeo)$ by a Gaussian $G(\phibar, 
V_{\phibar})$ for each value of \phibar 
\begin{equation}
R^{(2)}(\phibar, \phigeo) = I_{\phibar} G(\phibar, V_{\phibar})
\Delta_{\phigeo}  
\label{eq:imaging-10}
\end{equation}
and replacing $R^{(3)}$ in Eq.~\ref{eq:imaging-9} according to
Eq.~\ref{eq:imaging-6}
\begin{equation}
\log{L} = \sum _{ie} \log{ \left( \sum _j \frac{G_{ie,j}
    f_j}{\sin{\phigeo}} 
\right)} + \sum _{ie} \log{ \frac{ I_{\phibar} d\chi d\psi
    \cos{\psi}}{2\pi} } - \sum _j T_j f_j + C 
\label{eq:imaging-11}
\end{equation}
where $T_j = \sum _{id} R^{(3)} _{id,j}$ is the sensitivity to the
$j^{th}$ pixel in the object region and the second term is a constant
since the parameters $I_{\phibar}$, $\chi$, $\psi$ do not change for
each event. Eq.~\ref{eq:imaging-11} is therefore simplified as 
\begin{equation}
\log{L} = \sum _{ie} \log{ \left( \sum _j \frac{G_{ie,j}
    f_j}{\sin{\phigeo}} 
\right)} - \sum _j T_j f_j + C 
\label{eq:imaging-12}
\end{equation}
which is the likelihood function in list mode. We are also applying a
Newton-Raphson algorithm to maximize the list-mode likelihood function
in equation~\ref{eq:imaging-12}. 

\section{\label{sec:4} Imaging results }

\begin{figure}[htb]
\centering
\includegraphics[bbllx=70,bblly=505,bburx=330,bbury=730,
	width=0.5\linewidth,clip]{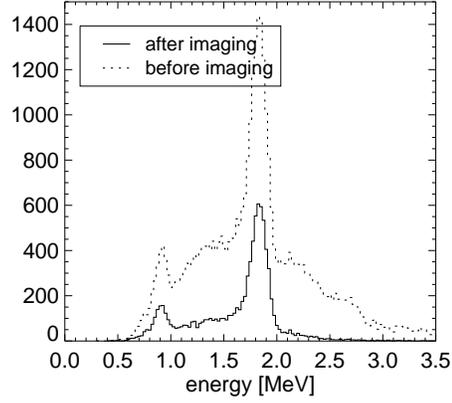}
\caption{\yt\ 3-site energy spectrum before and after software
  collimation.} 
\label{f:ytdi_4}
\end{figure}

\begin{figure}[htb]
\centering
\includegraphics[bbllx=55,bblly=405,bburx=550,bbury=730,
	width=.95\linewidth,clip]{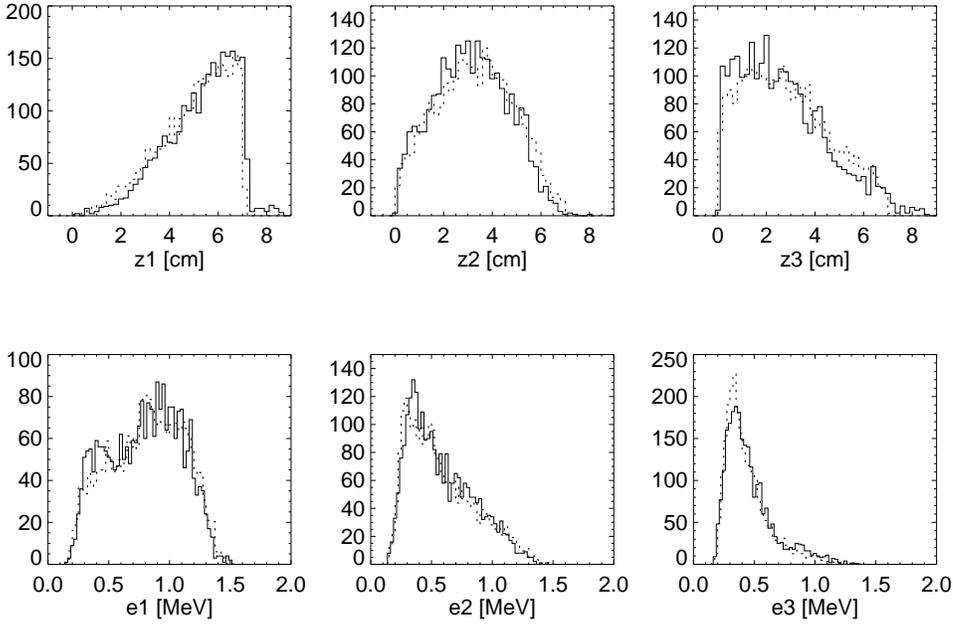}
\caption{Multi-site events selecting the 1.836~\MeV\ line and applying
  software collimation. {\it Top:} $z$ distribution for the first,
  second and third interaction. {\it Bottom:} energy spectra for the
  first, second and third interaction.}  
\label{f:ytdi_3}
\end{figure}

\begin{figure}[htb]
\centering
\includegraphics[width=0.60\linewidth,clip]{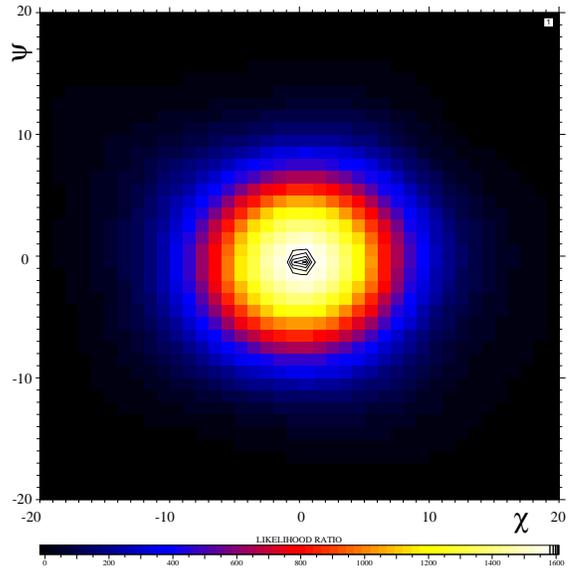}
\caption{ Maximum likelihood (list mode) image of an \yt\ source 2~m
  above the detector. }
\label{f:y88-image}
\end{figure}

The Maximum Likelihood imaging techniques in list mode has been used
to produce images of calibration \g-ray sources. Fig.~\ref{f:ytdi_4}
shows the energy spectrum obtained from exposing LXeGRIT to a 2738 kBq
\yt\ source at a distance of 2 m, on axis, for about 90 minutes;
3-site events have been selected. Before any selection, the main
features in the energy spectrum are the two \yt\ lines (0.898 and
1.836~\MeV), together with a continuum which extends up to
$\sim$3.7~\MeV\, due to partially absorbed \g-rays and, above
1.836~MeV, to pile-up of independent \g-rays. The energy spectrum
after selecting events in the ARM peak (a selection also called
software collimation) has been superimposed.   
The continuum is reduced by a factor of 4 at 1.5~\MeV\ and to 
a negligible fraction above 2~\MeV. The intensity of the 1.836~\MeV\  
line is reduced by 45\% by the ARM cut, consistent with the results 
presented in Sec.~\ref{sec:3}.   
The $z$ and energy distributions for each of the three interactions are
shown in Fig.~\ref{f:ytdi_3}, for events in the 1.836~\MeV\ full energy
peak and after software collimation. The same distributions from MC
data reproducing the experimental conditions have been
superimposed. The shape of the $z$ distributions are as expected for a
source on top of the detector, given that the first scatter is most
likely in the forward direction.     
The image of the source for the 1.836~\MeV\ line is shown in
Fig.~\ref{f:y88-image}, reconstructed with a list mode Newton-Raphson
algorithm. The source location is correctly determined with an
accuracy of about one pixel, i.e. 1\deg. 

A second example is the resolved image of two calibration sources,
\nuc{60}{Co} (1.17 and 1.33~\MeV) and \nuc{22}{Na} (1.27~\MeV). The
two sources were placed $\sim$1.7~m above the detector with angular
separation of $\sim$10\deg. A flat diffuse background and 100\%
detection efficiency $T_j$ for each pixel in the object region were
assumed, together with a variance of $V_{\phibar}$=3.5\deg.  

\begin{figure}[htb]
\centering
\includegraphics[width=0.60\linewidth,clip]{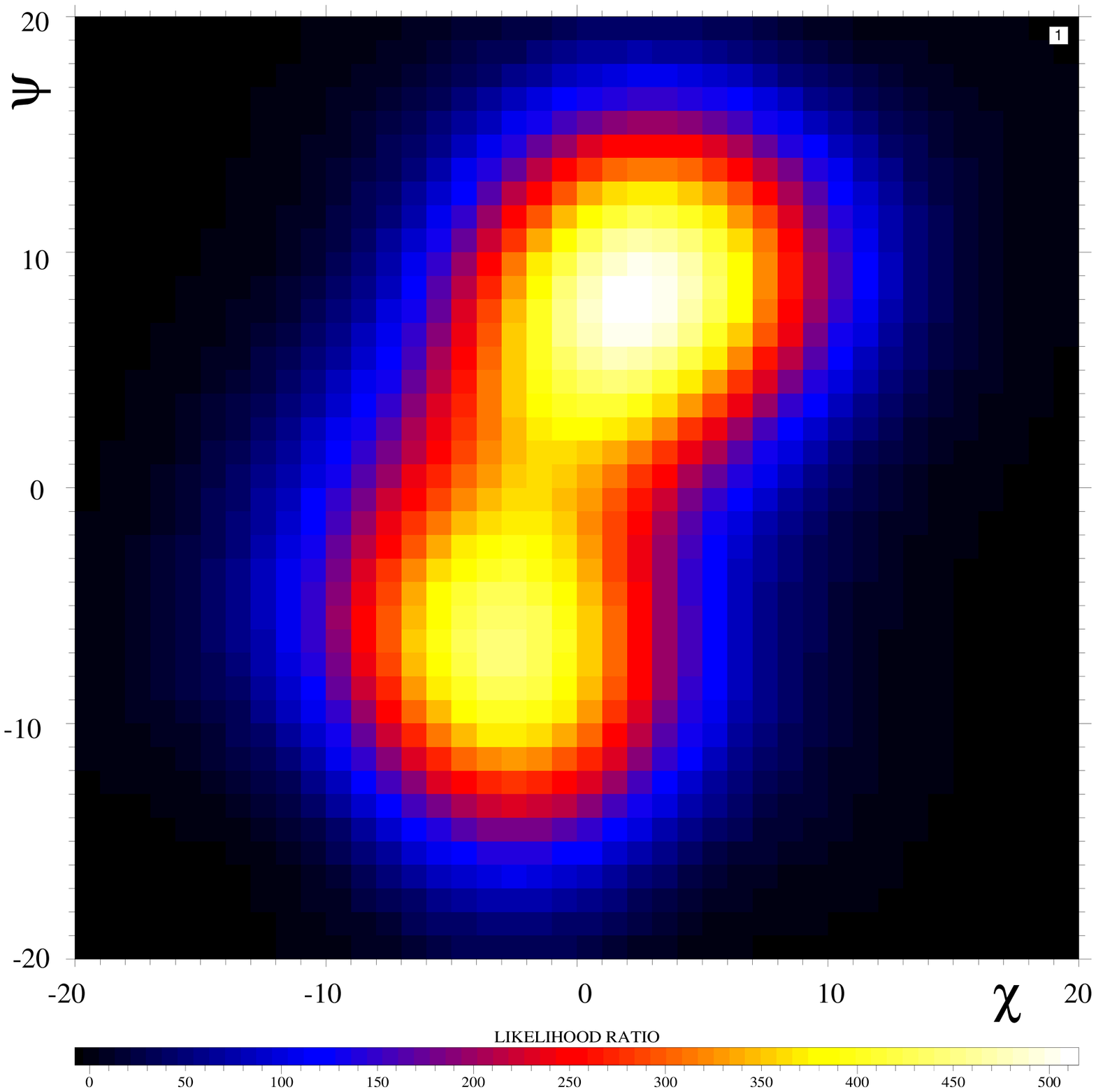}
\caption{ Maximum likelihood (list mode) resolved image of two
  calibration sources, \nuc{60}{Co} and \nuc{22}{Na}. } 
\label{f:na22+co60-image}
\end{figure}

\section*{Conclusions}

LXeGRIT is the first fully developed and tested prototype of Compton
Telescope based on a single position sensitive detector, such as a
LXeTPC with combined event energy and 3D localization in one large
homogeneous volume.
In this paper the details of its performance in imaging \MeV\ \g-ray
sources have been presented.
LXeGRIT has shown good performance as a \g-ray imager,
achieving an angular resolution of $\sim$4\deg\ at 1.8~\MeV,
consistent with expectations based on energy resolution, position
resolution and geometry of its TPC. Maximum Likelihood imaging
techniques have been successfully applied to the LXeGRIT data.   

\section*{Acknowledgments}

This work was supported by NASA grant NAG5-5108 to the Columbia 
Astrophysics Laboratory. S. Zhang acknowledges support by the Special
Founds for Major State Basic Research Projects and by the National
Natural Science Foundation of China via 10733010 and KJCX2-YW-T03. 


\end{document}